# An adaptive Origin-Destination flows cluster-detecting method to identify urban mobility trends

Mengyuan Fang, Luliang Tang, Zihan Kan, Xue Yang, Tao Pei, Qingquan Li, Chaokui Li

**Abstract:** Origin-Destination (OD) flow, as an abstract representation of the object`s movement or interaction, has been used to reveal the urban mobility and human-land interaction pattern. As an important spatial analysis approach, the clustering methods of point events have been extended to OD flows to identify the dominant trends and spatial structures of urban mobility. However, the existing methods for OD flow cluster-detecting are limited both in specific spatial scale and the uncertain result due to different parameters setting, which is difficult for complicated OD flows clustering under spatial heterogeneity. To address these limitations, in this paper, we proposed a novel OD flows cluster-detecting method based on the OPTICS algorithm which can identify OD flow clusters with various aggregation scales. The method can adaptively determine parameter value from the dataset without prior knowledge and artificial intervention. Experiments indicated that our method outperformed three state-of-the-art methods with more accurate and complete of clusters and less noise. As a case study, our method is applied to identify the potential routes for public transport service settings by detecting OD flow clusters within urban travel data.

## 1. Introduction

Geographical object`s movement and communication, such as population migration(Scardaccione, Scorza, Casas, & Murgante, 2010), merchandise trade(Bosnjak, Bilas, & Novak, 2019), and travel(Munizaga & Palma, 2012), can be modeled as Origin-Destination (OD) flow, which is regarded as a vector with spatial location(Y. Liu, Tong, & Liu, 2015). Compared with the trajectory model, OD flow ignores the specific movement path or trip mode but focuses on the travel purpose and behavior(He, Zhang, Chen, & Gu, 2018; Mao, Ji, & Liu, 2016; Toole et al., 2015).

With the fast growth of communication and mobile position technology, massive amounts of OD data with fine-granularity location are generated, making them have become an important data source for mining spatio-temporal patterns of urban mobility and analyzing the human-land interactions(Fang et al., 2021; Ferreira, Poco, Vo, Freire, & Silva, 2013; Tao & Thill, 2016b; Toole et al., 2015). By extending spatial analysis methods of point events to OD flows, researchers understood the pattern of urban mobility and human-land interaction. Clustering, as an important and commonly used spatial analysis method, has been widely applied to OD flow data to reveal dominant trends and spatial structure of urban mobility. To detect OD flow clusters, Xi Zhu and Guo (2014), Tao and Thill (2016b), Tao, Thill, Depken and Kashiha (2017), C. Song, Pei, and Shu (2020) proposed the cluster-detecting methods of OD flows based on the density-based, spatial statistical-based, hierarchical-based clustering methods respectively. However, these existing methods are hard to be applied to complicated OD flows in the real world. Firstly, the aggregation scale of OD clusters is various due to the spatial heterogeneity(Shu et al., 2020), yet most of the existing methods can only identify the clusters with a specific scale under a global parameter setting. Secondly, the parameter selection process is complicated and lacking objective criteria; thus, the result will be various and uncertain under parameter values set by different experiences (Ankerst, Breunig, Kriegel, & Sander, 1999).

Aiming to address the above issues, in this study, we proposed a novel OD flows cluster-detecting method based on OPTICS (Ordering Points To Identify the Clustering Structure) algorithm. Firstly, without the need for manual intervention, the method adaptively acquires the parameter setting based on Ripley's K function (Ripley, 1976) and Kullback-Leibler Divergence (Leibler, 1951). The K function obtains the range of aggregation scale and ensures the statistical significance of the clusters to avoid generating false clusters. Secondly, the clusters mixed in OD flows with various aggregation scales are identified using the OPTICS algorithm. As an advanced density-based clustering method, the OPTICS can adaptively identify clusters with various local densities by identifying the density structure of the dataset. Finally, the noise mixed in clusters is filtered out using the Local Outliers Factor (Breunig, Kriegel, Ng, & Sander, 2000), making the result depict mobility trends of clusters accurately. In summary, our method can detect OD flow clusters adaptively and accurately, which can be used to mine "regular routes" or "collective routes" among urban mobility data under spatial heterogeneity.

The contributions of this paper can be summarized as follows:

1) A novel OD flows cluster-detecting method is proposed based on the OPTICS algorithm, which is proven to outperform the three state-of-the-art methods in identifying mixed clusters with various aggregation scales.

2) A parameters selection strategy is proposed based on spatial statistics and information theory, which can adaptively acquire the parameter setting of the OPTICS algorithm without prior knowledge and manual intervention.

3) The proposed method is applied to detect commuter OD flow clusters as a case study, whose results provide effective support for urban public transport service settings.

The remainder of this paper is organized as follows: Section 2 summarizes related work on spatial analysis using OD flow data. Section 3 introduces the proposed method. Section 4 evaluates the method with simulative experiments qualitatively and quantitatively. Section 5 presents a case study using the urban mobility OD flow data to identify commuter-group routes. Finally, Section 6 concludes the work.

## 2. Related Works

Using OD flow data, existing research studied the spatio-temporal pattern of urban population activities and the human-land interaction. According to the research object, these researches can be divided into region-oriented ones and individual-oriented ones.

The region-oriented research mainly focused on cross-regional spatial interaction analysis and regional mobility characteristics using travel or communication OD flow data; thus, OD flow data are merged into regions. To analyze the spatial interaction, Paul Lesage and Polasek (2008), Tsutsumi and Tamesue (2012), Thompson, Saxberg, Lega, Tong, and Brown (2019) respectively proposed spatial econometric flow based on the gravity model to reveal the relative connectivity between regions under commodity trade, migration, and transportation. Griffith and Jones (1980), Marrocu and Paci (2013), Grunfelder, Nielsen, and Groth (2015) explored the relationship between spatial interaction and factors such as urban structure, economics, and environment. Ujiie and Fukumoto (2012), X. Liu, Gong, Gong, and Liu (2015), Tang et al. (2019) modeled intra-city spatial interaction and further detected community to reveal functional region and community structures. Abel and Sander (2014), Ci Song et al. (2019) interpreted the trends, patterns of OD flows between regions. To understand the regional mobility characteristics, Xu et al. (2015); Xu et al. (2016) considered daily activity range, number of activities, and movement frequency to reveal residents` mobility patterns in different urban regions. Martins, da Silva, and Pinto (2019) assessed resilience of transportation systems using OD flow data. Yang, Sun, Shang, Wang, and Zhu (2019) delimited seven mobility patterns of road intersection regions according to the O/D time series and illustrated that the regional functions have apparent effects on these mobility patterns. However, these region-level analyses are difficult to accurately detect behavior patterns of the population due to the MAUP (Wong, 2009) and the coarse spatial granularity exerted by regional division(Tao & Thill, 2016a).

Apart from the regional-oriented research, the individual-oriented is dedicated specifically to analyzing spatial heterogeneity under a fine granularity. Taking individual as the research object, researchers analyzed OD flow data by defining the proximity measure of OD flows (Shu et al., 2020; Tao & Thill, 2016a) and extending spatial analysis methods of point events to OD flows, such as spatial autocorrelation(Black, 1992; Y. Liu et al., 2015), spatial association (Berglund & Karlström, 1999; Tao & Thill, 2020), Ripley`s K-function (Tao & Thill, 2016b), L-function(Shu et al., 2020), spatial scan statistics (Gao, Li, Wang, Jeong, & Soltani, 2018) and clustering methods(X. Guo, Xu, Zhang, Lu, & Zhang, 2020; C. Song et al., 2020; Tao et al., 2017) to analyze the spatial heterogeneity and pattern among OD flows. As an important approach, clustering can identify the dominant trends and spatial structures at the individual level (Xiang & Wu, 2019; Yao et al., 2018). D. Guo and Zhu (2014), Tao and Thill (2016b), Tao et al. (2017), X. Zhu, Guo, Koylu, and Chen (2019), C. Song et al. (2020) respectively proposed several clustering methods for OD flows, based on density-based, spatial statistical-based, hierarchical-based methods. These methods effectively detect the "hot flow" of theft and recovery location pairs, travel trends, and spatial structure of the commodity transaction networks. Considering temporal dimension, Faroqi, Mesbah, and Kim (2017), Yao et al. (2018), Xiang and Wu (2019) identified significant spatio-temporal trends within OD flow data. Although the above methods have been applied to reveal mobility patterns at the individual level, these existing methods are hard to be used to complicated OD flows under spatial heterogeneity. Due to the spatial heterogeneity, clusters with various aggregation scales are mixed within OD flow data(Shu et al., 2020). However, under a specific parameter setting, most of the existing methods can only detect clusters at the specific aggregation scale, yet the clusters with a larger difference of scale will be merged incorrectly or neglected. Besides, the results will be uncertain due to the parameter set by different prior experiences and manual intervention (Ankerst et al., 1999).

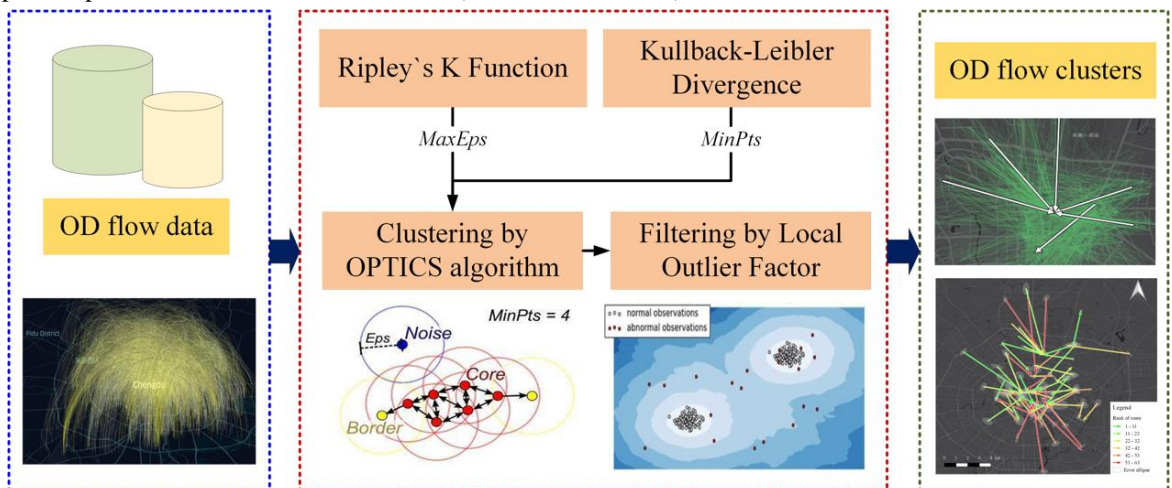

Fig.1. Flowchart of the proposed method

## 3. Methodology

In this section, a novel OD flow cluster-detecting method is proposed based on the OPTICS algorithm. Fig.1 shows the flowchart of the method. Firstly, the method adaptively acquires the parameter setting of the OPTICS algorithm based on Ripley's K function and Kullback-Leibler Divergence without prior experience and manual intervention. Secondly, the clusters in OD flows with various aggregation scales are identified using the OPTICS algorithm. Finally, the noise mixed in clusters is denoised using the Local Outlier Factor.

### 3.1 Definitions about OD flow

Before introducing the specific method, the basic concepts of OD flows are defined, containing the definition, distance, and spatial randomness of OD flows used in clustering. Fig. 2 demonstrates the basic model of OD flows.

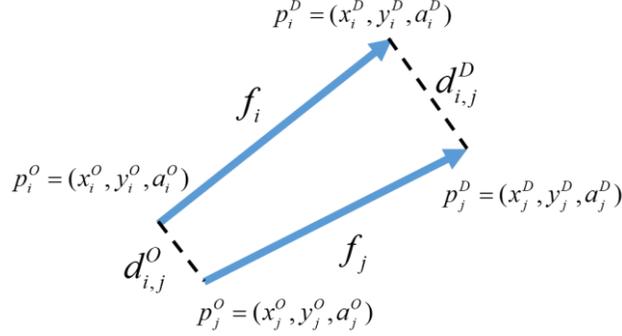

Fig. 2. OD flow model

**Definition 1 (OD Flow)**: An OD flow is defined as a spatial vector representing individual movement or spatial interaction whose endpoints have geographic coordinates. The OD flow $i$ can be expressed as $f_i = <p_i^O, p_i^D>$ where the origin-point $p_i^O = (x_i^O, y_i^O, a_i^O)$ and the destination-point $p_i^D = (x_i^D, y_i^D, a_i^D)$ consist of their coordinates and attributes (such as time).

**Definition 2 (Distance of OD flows)**: The distance of OD flows is the fundamental and indispensable measurement in the clustering, which can be defined in terms of both $d^O$ and $d^D$. In general, two OD flows whose Origin and Destination points are close to each other have a smaller distance of OD flows.

According to the existing researches, three types of OD flows` distance are defined as (1-3):

*Chebyshev Distance(C. Song et al., 2020)*:

$$\rho^c(f_i, f_j) = \max(d_{i,j}^O, d_{i,j}^D) \tag{1}$$

*Manhattan Distance(Shu et al., 2020)*:

$$\rho^m(f_i, f_j) = d_{i,j}^O + d_{i,j}^D \tag{2}$$

*Flow distance(Tao & Thill, 2016b)*:

$$\rho^f(f_i, f_j) = \sqrt{\frac{a(d_{i,j}^O)^2 + b(d_{i,j}^D)^2}{L_i \cdot L_j}} \tag{3}$$

Where the $a$ and $b$ represent the weight of $d^O$ and $d^D$ respectively; the $L_i$ and $L_j$ represent the length of the $f_i$ and $f_j$ respectively.

**Definition 3 (Spatial randomness of OD flows)**: Complete Spatial Randomness (CSR) describes events occurring within the study area in a completely random pattern. It is commonly used for generating the null distribution to test the spatial homogeneity(Ripley, 1976). Similarly, the spatial randomness of OD flows is defined as the spatial random process of flows to test the spatial homogeneity of the OD flows. The CSR of OD flows can be generated by randomly linking two sets of randomly distributed points. By taking the length, direction, and distribution of flows into account, conditional spatial randomness of OD flows can also be generated by shifting or recombining OD flows` endpoints (Y. Liu et al., 2015).

### 3.2 Determining the parameters adaptively by understanding the OD flows

According to the local density, OD flows can be divided into dense flows or sparse flows(C. Song et al., 2020). The former form clusters, while the latter are random and regarded as noise. Identifying clusters and noise by local density, density-based clustering methods are widely used in spatial cluster detection. The key idea of density-based clustering is to identify the core-objects with at least a minimum number of objects (MinPts) within a given aggregation scale (Eps). Further, the core-objects form clusters with all objects that are reachable from it, and the other objects are labeled as noise (Ankerst et al., 1999). Without prior knowledge about the dataset, determining optimal parameter value is always a problem (Song et al. 2020).

To setting the parameters adaptively, we understand the OD flows using spatial statistics and information entropy theory.

### 3.2.1 Obtaining the optimal MaxEps value by Ripley`s K function

Existing researches used spatial statistics methods, such as K-function, L-function, to test OD flows aggregation pattern under the specific spatial scale (Shu et al., 2020); however, the testing result is hard to support the cluster-detecting since their clustering methods can only detect the clusters with the specific aggregation scale. In this research, we adopt the OPTICS algorithm which can adaptively identify the clusters with a range of scale. In the OPTICS, the scale parameter of density-based clustering, Eps, is replaced with a threshold of scale, MaxEps, and the cluster whose aggregation scale is below the threshold can be identified. To set the parameter legitimately, we adopt Ripley`s K-function to obtain the maximum aggregation scales of OD flows with a significant clustered pattern.

Ripley`s K function, as a representative spatial statistical method, is widely used to determine whether the OD flows appear to be dispersed, clustered, or randomly distributed at a given spatial scale. The K function is defined as (4). To be specific, the K function counts the number of neighboring OD flows within a given distance of each OD flow, then compares the result with the confidence interval resulting from the spatial random dataset to test the randomness hypothesis (Shu et al., 2020; Tao & Thill, 2016b). As Fig. 3 demonstrates, when the K function`s value is higher than the upper bound of the confidence interval, the OD flows appear clustered pattern at that scale. Therefore, the parameter MaxEps is set as the upper bound of the clustered scale.

$$K(r) = \lambda_F^{-1} \frac{\sum_{i=1}^{n} \sum_{j=1}^{n} \sigma_{ij}^{F}(r)}{n} \ (i \neq j) \qquad (4)$$

Where $n$ is the number of OD flows; $r$ is the scale being analyzed; $\sigma_{ij}^{F}(r)$ is set as 1 or 0 depending on whether the distance between $f_i$ and $f_j$ is shorter than $r$; $\lambda_F$ represents the density of the flow dataset which can be estimated by referring to (Shu et al., 2020).

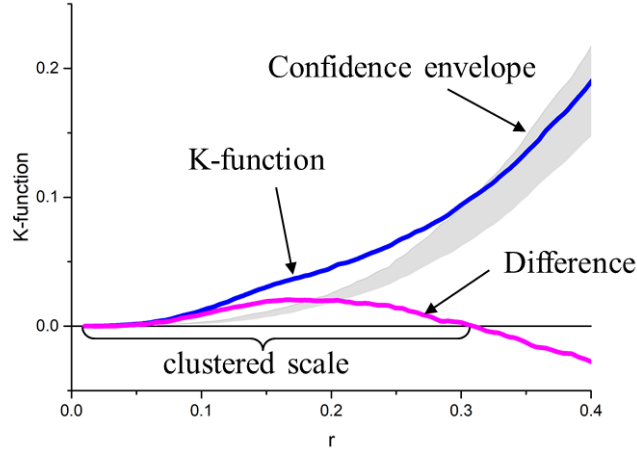

Fig. 3. Diagram of K function testing (Gray belts are the confidence envelope of randomness hypothesis; Difference represents the difference between the function and the upper bound of the confidence envelope)

Since the O/D points follow a specific distribution pattern, such as located on both side of the roads (Gao et al., 2018), concentrated nearby the downtown district(Y. Liu et al., 2015), consistent with the residents`spatial distribution(Shu et al., 2020), comparing the K-function with that of completely spatial random OD flows are not rigorous enough(Tao & Thill, 2016b). Therefore, in this step, the conditional spatial random OD flow datasets generated by recombining the OD points of the measured dataset are adopted again by Monte Carlo simulation to generate the confidence envelope of randomness distribution(Y. Liu et al., 2015).

### 3.2.2 Obtaining the optimal MinPts value by Kullback-Leibler Divergence

The parameter MinPts describes the number of flows required to form a cluster within neighborhood distance (the aggregation scale). It is crucial in the cluster-detecting which directly affects the results: If the MinPts value is too large, a mass of clustered flows will be misclassified as noise; on the contrary, the noise may be mistakenly generated as clusters. Ci Song et al. (2019) generated the histogram of the kth nearest flow distances and adopted the k making the histogram most separable as the MinPts value. However, under spatial heterogeneity, lots of clusters of various sizes are mixed in the OD flows; thus, the histogram is too confused to identify the mixed Poisson distribution by the Density Domain model. To obtain the MinPts value making the best distinction of OD flow clusters and noise flows, in this study, we use the Kullback-Leibler Divergence(KLD) to measure the information within the histogram under different k value.

In information theory, the KLD, also called relative entropy, measures how much one probability distribution is different from a reference probability distribution. As shown in (5), in our method, KLD is adopted to evaluate the difference of the kth nearest flow distances` histogram between the dataset and the spatial random OD flow dataset. The larger the KLD, the more the information contained compared with spatial random dataset under that value of k; in other words, the clusters within OD flows are more distinguishable. Therefore, when the KLD reaches its maximum value, that value of k is adopted as the MinPts value in cluster-detecting.

$$D_{KL}(p^k \| q^k) = \sum_{i=1}^{N} p^k(x_i) \times \log \frac{p^k(x_i)}{q^k(x_i)}$$
(5)

Where $p^k$ and $q^k$ represent the kth nearest flow distances` histogram of the original OD flow dataset and that of spatial random OD flow dataset, respectively. And the construction of the histogram can refer to (C. Song et al., 2020).

In this step, the conditional spatial random OD flow dataset is again adopted which is generated by recombining the O/D points of the original dataset as Section 3.2.1. Besides, we simulate the random dataset multiple times and construct the histogram of spatial random OD flow dataset by averaging each result`s histogram to improve the reliability of the MinPts obtaining.

### 3.3 OD flow cluster-detecting using the OPTICS algorithm

The density-based clustering methods such as DBSCAN (Density-Based Spatial Clustering of Applications with Noise) were widely used in spatial cluster detection, which can find arbitrarily shaped clusters and identify the noise. However, as is mentioned above, the methods cannot be applicable to the clusters with large scale difference: if the aggregation scale of one cluster is larger than the given scale, the cluster will be regarded as noise; on the contrary, the clusters will be mixed with noise flow or the adjacent clusters may be incorrectly merged. Therefore, in this research, we adopt an improved method of density-based clustering, the OPTICS algorithm, to identify clusters with various scales adaptively.

The OPTICS algorithm generates the clustering structure of the dataset which supports identifying clusters adaptively. The generated clustering structure is presented as Reachability Plots, as is shown in Fig. 4, where the clusters show up as valleys, and the deeper the valley, the denser the cluster. Clusters are identified by detecting the steep upward and downward of the valleys. Meanwhile, the reachability-plot`s appearance will not change significantly under parameter values around the optimal value; thus, the OPTICS algorithm has better robustness (Ankerst et al., 1999).

After the cluster-detecting, there is still some noise mixed in clusters which interferes with the analysis of clustering results. Thus, we further filter the noise by Local Outlier Factor, referring to (Breunig et al., 2000).

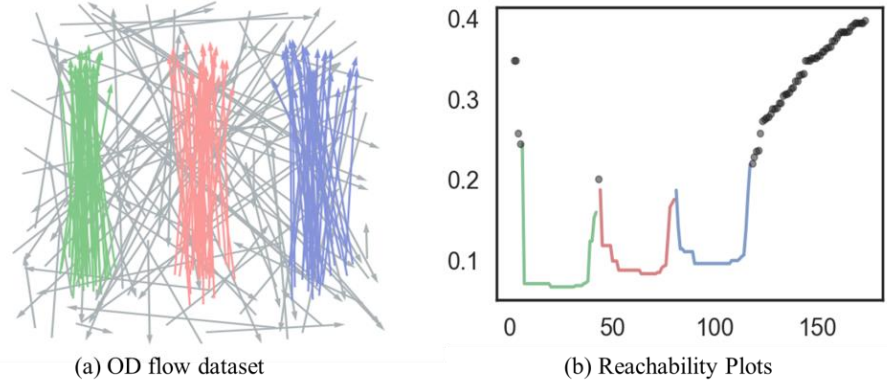

(a) OD flow dataset        (b) Reachability Plots

Fig. 4. Flow data and its reachability plots (The colors in (b) correspond to that in (a), and black dots in (b) represent noise)

## 4. Simulation Experiment

In this section, we experiment with the simulated dataset and analyze the result of cluster-detecting with the predefined ground-truth label to evaluate the effectiveness and superiority of the proposed method. The result was qualitatively and quantitatively compared with three state-of-the-art methods as benchmarks.

### 4.1 Simulated flow datasets

This experiment simulated OD flow datasets containing both clustered and random OD flows. To adapt to the spatial heterogeneity, the aggregation scale of each cluster was various, and clustered OD flows is less than random ones. Further, two simulated flow datasets with different cluster sizes were constructed for comparison. As Fig. 5 shown, two simulated datasets shared similar spatial distribution, structure, and the same total quantity of OD flows. There were about 10 OD flows in a single cluster in dataset A and 30 in dataset B.

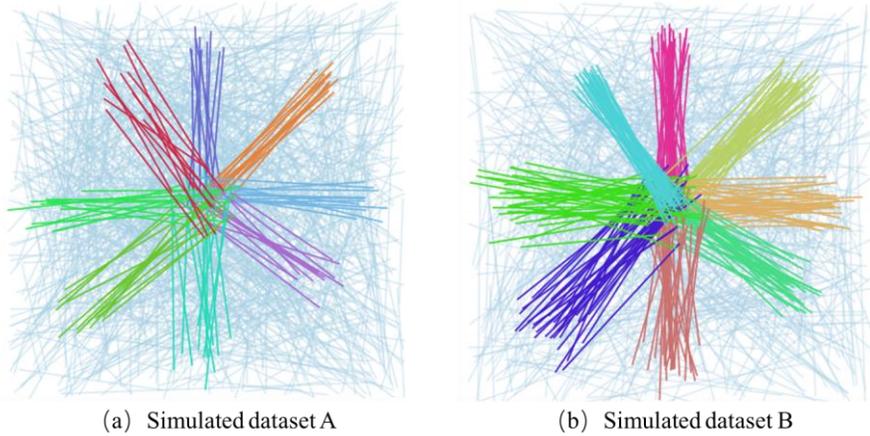

(a) Simulated dataset A　　　　(b) Simulated dataset B

Fig. 5 Simulated OD flow datasets with different cluster sizes. Each color represents an OD flow cluster, and random OD flow is set as gray.

## 4.2 Cluster detection and comparison

### 4.2.1 Parameters setting adaptively

The proposed method is implemented using Python 3.8 and Sklearn library (Pedregosa et al., 2012). Firstly, as shown in Fig. 6, the two datasets are understood by K-function and KLD to setting parameter adaptively. In the experiment, the Monte Carlo simulation was implemented 99 times by generating conditional spatial random OD flows to obtain the confidence envelopes with a 99% significance level.

The aggregation scale range of the two datasets are [0.02,0.16] and [0.015,0.175]. These two ranges are similar because the two datasets share a similar spatial structure, and the subtle difference is caused by accumulative effects (Kiskowski, Hancock, & Kenworthy, 2009). Meanwhile, the curves of KLD reach their maximum at the k values of 11 and 25, matching the two dataset's cluster sizes.

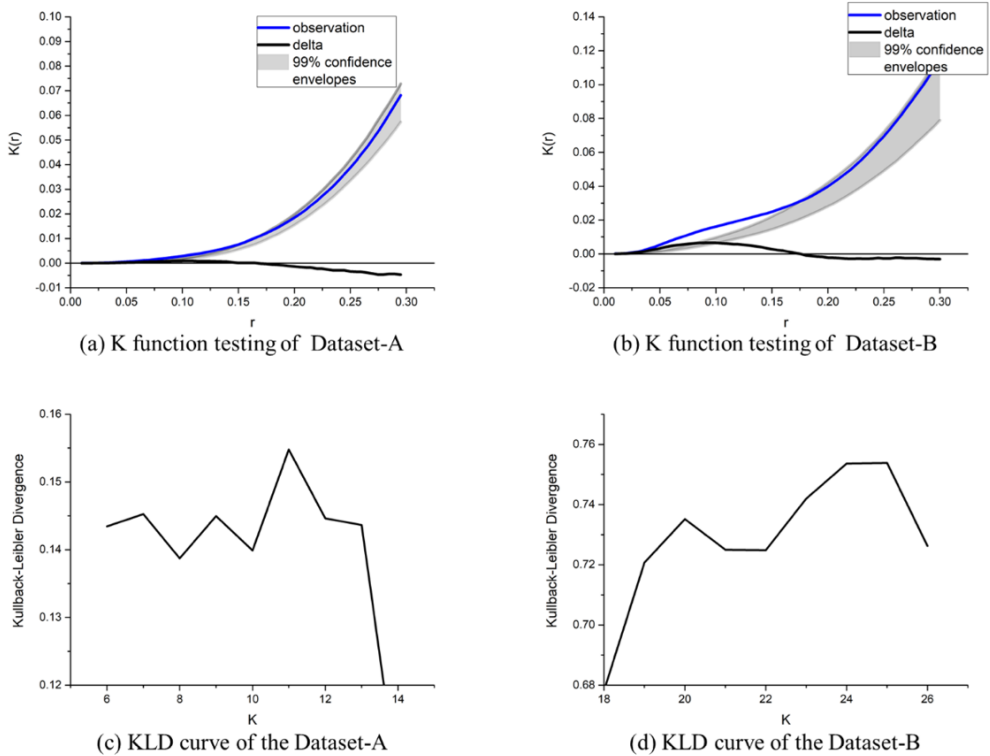

(a) K function testing of Dataset-A　　　　(b) K function testing of Dataset-B

(c) KLD curve of the Dataset-A　　　　(d) KLD curve of the Dataset-B

Fig. 6. Determination of the parameters

### 4.2.2 Benchmark methods and evaluation metric

To evaluate the results, we adopted three state-of-the-art methods of OD flow clustering, flowHDBSCAN(Tao et al., 2017), Local K-function(Tao & Thill, 2016b), and Density Domain Decomposition (C. Song et al., 2020) as benchmark methods. The specific methods and experimental parameter settings are determined according to the reference and shown in Table 1. Lack of a standard strategy for parameter selection, we selected different parameter values for the flowHDBSCAN method.

Silhouette Coefficient (Rousseeuw, 1987), V-measure (Rosenberg & Hirschberg, 2007), and Fowlkes Mallows index were adopted as metrics to evaluate the results quantitatively. The definitions for these three indicators are as follows.

**Definition 3 (Silhouette Coefficient)**: Silhouette Coefficient, calculated as (6), is a widely used metric in clustering to measure how an object is similar to its cluster compared to other clusters. It provides an evaluation of clustering validity free of truth-label. The value of the Silhouette Coefficient ranges from -1 to 1, and the larger the value, the higher the clustering effectiveness.

$$SC = mean(\frac{b_i - a_i}{\max(a_i, b_i)})(i = 1, 2, ..., n) \tag{6}$$

Where $a_i$ represents the mean distance between object $i$ and all other objects in the same cluster, and $b_i$ represents the mean distance between object $i$ and all other objects in the next nearest cluster.

**Definition 4 (V-measure)**: V-measure is an entropy-based measure that explicitly measures how successfully the cluster-detecting result satisfied homogeneity and completeness criteria. The V-measure value is also between 0-1, and a higher value indicates a better clustering result. For more details, refer to (Rosenberg & Hirschberg, 2007).

**Definition 5 (Fowlkes Mallows Index)**: Fowlkes Mallows Index, calculated as (7), is the geometric mean of precision and recall, which measures how successfully both the two criteria have been satisfied. The Fowlkes Mallows Index`s value is between 0-1, and a higher value indicates a greater similarity between the clusters and the benchmark classifications, namely, a better clustering result.

$$FMI = \frac{TP}{\sqrt{(TP + FP)(TP + FN)}} \tag{7}$$

Where $TP$ is the number of true-positive objects, $FP$ is the number of false-positive objects, and $FN$ is the number of false-negative objects.

**Table 1**
Experimental methods and parameter settings

| Method | Distance Measurement | Clustering algorithm | Parameter Setting | |
| --- | --- | --- | --- | --- |
| | | | Dataset-A | Dataset-B |
| flowHDBSCAN | Flow distance | Hierarchical-based | min_cluster_size=8/10/12 | min_cluster_size=10/15/20 |
| Local K function | Flow distance | Statistical-based | R=0.12 | R=0.12 |
| Density Domain Decomposition | Chebyshev distance | Density-based | MinPts =11,Eps=0.135 | MinPts =12,Eps=0.145 |
| our method | Chebyshev distance | Density-based | Minpts=11, MaxEps=0.16 | Minpts=25,MaxEps=0.175 |

### 4.2.3 Results and analyses

Two simulated datasets were processed by our method and benchmark methods. The results are visualized in Fig. 7, in which flows in each color represent each detected cluster, and the quantitative comparison are shown in Table 2.

According to the result, the cluster-detecting of the benchmark methods have below shortcomings: 1) although the Local K function method achieved the highest Silhouette Coefficient, due to the limitation of the aggregation scale, the method ignored some clusters with lower density and some clusters are not identified integrally; 2) some clusters identified by the Density Domain Decomposition method were mistakenly merged; 3) since it apparently changes under similar parameter value, the result of the flowHDBSCAN method is sensitive to the parameter value, then, fake clusters were generated under improper parameter setting.

Compared with the three benchmark methods, our method achieved a better cluster-detecting result with more complete identification, less noise mixed, and better quantitative evaluation. The advantage of our method can be summarized as below: 1) generating the cluster structure by OPTICS made the identified clusters with various aggregation scales and avoided the clusters being merged or ignored incorrectly; 2) the K function ensured the statistical significance of the clusters to avoid generating false clusters; 3) the LOF method denoised the results that made the depiction of mobility trends of clusters more accurate.

In summary, our method outperforms the three benchmark methods in detecting OD flow clusters with various densities and has higher accuracy and validity.

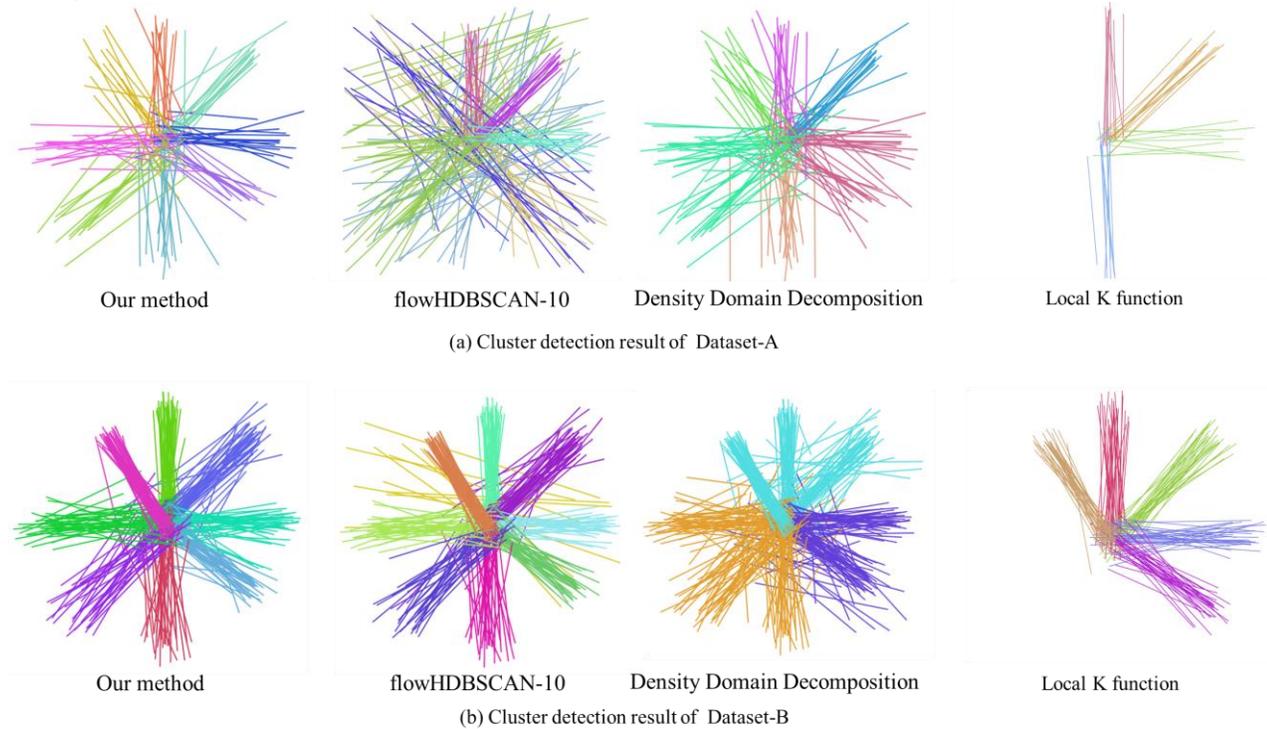

(a) Cluster detection result of Dataset-A

(b) Cluster detection result of Dataset-B

**Fig. 7.** Qualitative comparison of cluster-detecting results

**Table 2**
Quantitative comparison of the cluster-detecting result

| Data | Method | | clusters | V measure | Silhouette Coefficient | Fowlkes-Mallows |
|------|--------|--|----------|-----------|------------------------|------------------|
| Dataset-A | Our method | | **8** | **0.79** | 0.58 | **0.96** |
| | flowHDBSCAN | Mcs=8 | 12 | 0.51 | 0.44 | 0.82 |
| | | Mcs=10 | 7 | 0.39 | 0.44 | 0.8 |
| | | Mcs=12 | 3 | 0.23 | 0.01 | 0.73 |
| | Density Domain Decomposition | | 6 | 0.7 | 0.49 | 0.94 |
| | Local K function | | 4 | 0.55 | **0.71** | 0.92 |
| Dataset-B | Our method | | **8** | **0.87** | 0.64 | **0.94** |
| | flowHDBSCAN | Mcs=10 | 9 | 0.86 | 0.67 | 0.93 |
| | | Mcs=15 | 7 | 0.85 | 0.64 | **0.94** |
| | | Mcs=20 | 6 | 0.81 | 0.54 | 0.91 |
| | Density Domain Decomposition | | 3 | 0.64 | 0.49 | 0.84 |
| | Local K function | | 5 | 0.71 | **0.69** | 0.84 |

## 5. Case Study: the potential routes of public transportation service

Commuting refers to regular and repeated traveling between locations, especially the traveling from a residential area to work or study location(Wikipedia, 2020). Massive private commuting is a major contribution to traffic congestion and pollution; hence, commuters are encouraged to use public transportation or carpooling to reduce traffic pollution and save cost (Higgins, Sweet, & Kanaroglou, 2018).

To meet the travel demands of commuters effectively, the primary stage of public transportation service designing is to investigate routes of commuter groups and evaluate the potential of setting public transportation service on each route(Kong et al., 2018). However, the traditional traffic demand collection by on-line or questionnaire survey is costly, time-consuming, and hard to acquire the information completely and accurately. Therefore, in this study, we detected OD flow clusters in real-world urban travel data to identify routes of commuter groups using our method, and further, we evaluated the potential of public transportation service by analyzing the clusters.

## 5.1 Data description

We adopted the urban travel dataset provided by DIDI, which is a considerable part of transportation services in China (Wang et al., 2020). In the dataset, each record represents one trip including the latitude, longitude and time of origin and destination of the trip; thus, each record can be modeled as an OD flow. In this study, the urban area of Chengdu, China within the Ring Expressway is chosen as the study area and the travel data departing from 7:30 am to 8:00 am from November 7th to 11th, 2016 (5 weekdays) is extracted to analyze the commuting trips, which contains 21169 records.

Fig. 8 shows the spatial distribution of the OD flows and their origin and destination points used in this study. According to the visualization, the origins, namely the residences, are relatively dispersed while the destinations are relatively concentrated in the city center.

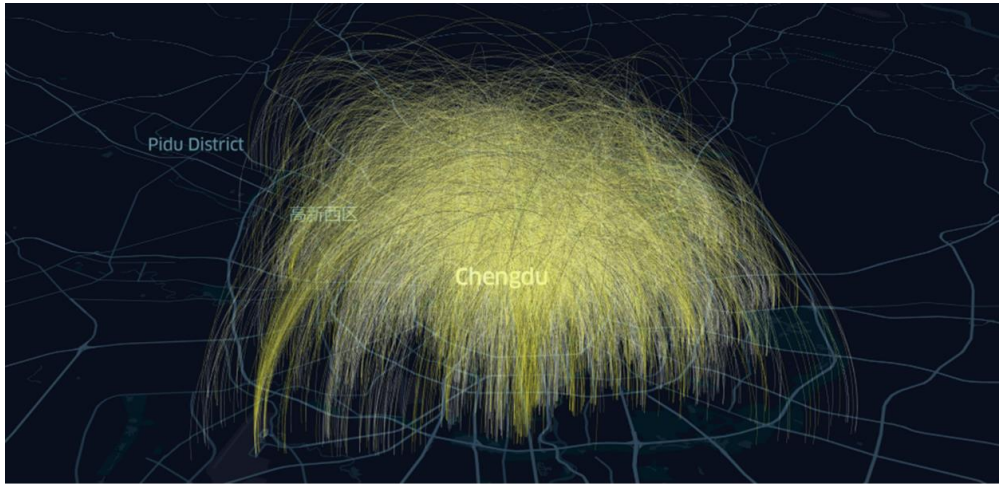

(a) Distribution of OD flow data in this study

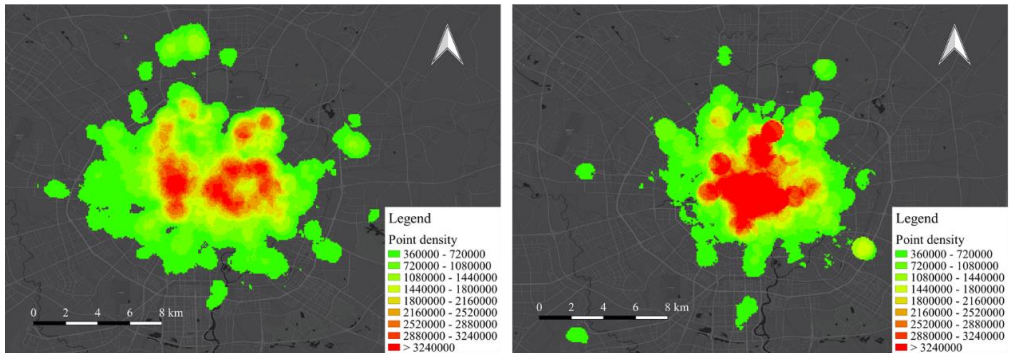

(b) Distribution of origin points          (c) Distribution of destination points

Fig. 8. visualization of the experimental dataset

## 5.2 Identifying routes of commuter groups by OD flow cluster-detecting

Before the cluster-detecting, the spatial coordinates were normalized to range 0-1 respectively to facilitate the calculation, and the Chebyshev Distance was adopted as the distance measurement of OD flows. Fig. 9 shows the result of K function testing and the KLD curve of the trip dataset. As Fig. 9 shown, the OD flows are significantly clustered within the scale range of [0,0.022] at the 99% significance level, and the KLD curve reaches its peak at k=14.

Then, the OPTICS algorithm detected the clusters with the values of 0.022 for MaxEps and 14 for MinPts. Fig. 10 shows the detecting result, which contains 63 clusters consisting of 1702 clustered trips. The cluster size ranges from 14 to 170, and the trip distance ranges from 1.3 to 8.4 kilometers. The distribution of OD points in each cluster is marked by the error ellipse. From the result, most of the trips depart from the residential area and travel towards the business district, metro stations, and hospitals.

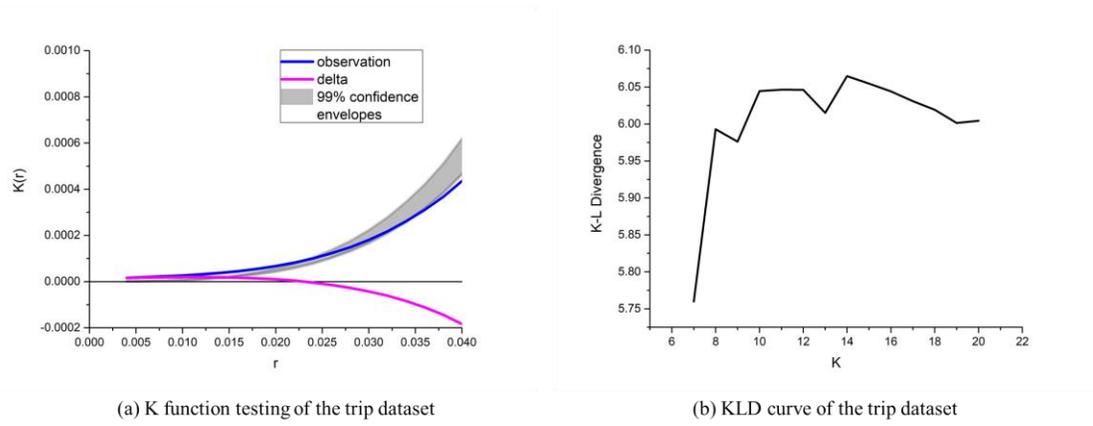

(a) K function testing of the trip dataset    (b) KLD curve of the trip dataset

Fig. 9. Determination of the cluster-detecting parameters

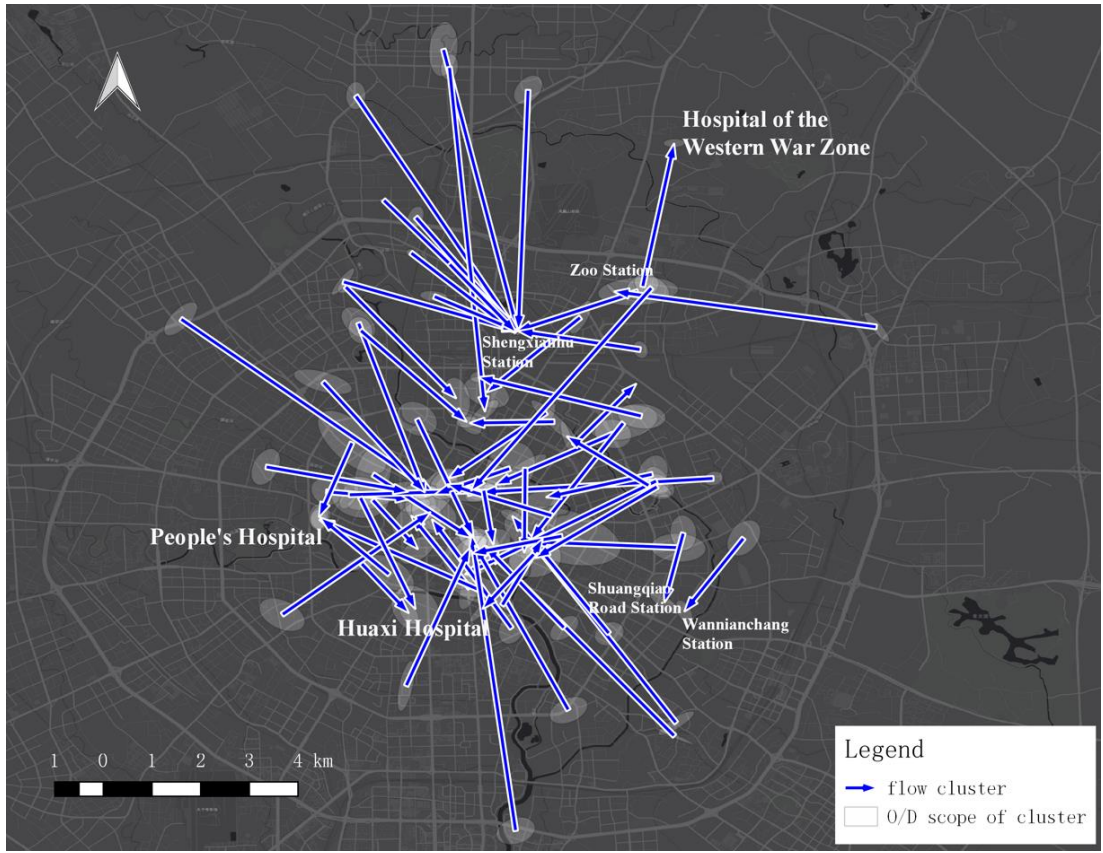

Fig. 10. The routes of commuter groups (OD flow clusters) identified from urban travel data

### 5.3 Analyzing the potential for setting public transportation service

Though the routes of commuter groups are obtained, not all the routes are suitable for setting public transportation service, such as temporary group traveled route, large O/D scope but sparsely traveled route. To evaluate the potential for setting public transportation service, in this study, we defined three indicators from three aspects of commuter group density, daily stability, and the concentration of departure times.

#### 5.3.1 Commuter Group Density

To evaluate the spatial concentration of each commuting group, *CGD*, is defined as the ratio of the cluster size to the product of its O/D error ellipse area, as (8) shown. The denominator is taken logarithm to balance the order of magnitude. Cluster with a higher *CGD* is more potential for setting public transportation service on the corresponding commuting route.

$$CGD_i = \frac{count_i}{\ln(Area_{Oi} * Area_{Di})} \quad (8)$$

Where $i$ is the index of a cluster; $count$ and $Area$ respectively represent the number of OD flow in the cluster and the area of O/D scope of the cluster.

The top 10 clusters of CGD are shown in Fig. 11 and Table 3. According to the result, the routes with high CGD are mainly concentrated in the center of the city and from business area to business area, such as the routes from Chengdu 339 to Kowloon plaza, from Zhongchuan Building to Huaxi International. This kind of cluster has shorter travel distances, more commuter volume, and a wider O/D scope. These clusters have a large aggregated scale, which is consistent with (Shu et al., 2020). However, despite a relatively high commuter volume, it is difficult to determine an optimal location as the O/D station for setting service station because the O/D scope location is so large. Besides, the clusters from the residential area to the subway station, such as from Shallow peninsula to Shengxianhu MTR station, from Huamanting to Shengxianhu MTR station, have a large commuter volume with concentrated O/D positions; thus, this kind of clusters are ideal for setting public transportation service. Meanwhile, there are also some routes with high CGD from resident area to hospital, such as from Shallow peninsula to City No.2 Hospital, from Confucian temple west street to Provincial People`s Hospital. These clusters are probably caused by the regional characteristics of health care services or the residences of medical staff, which require further analysis.

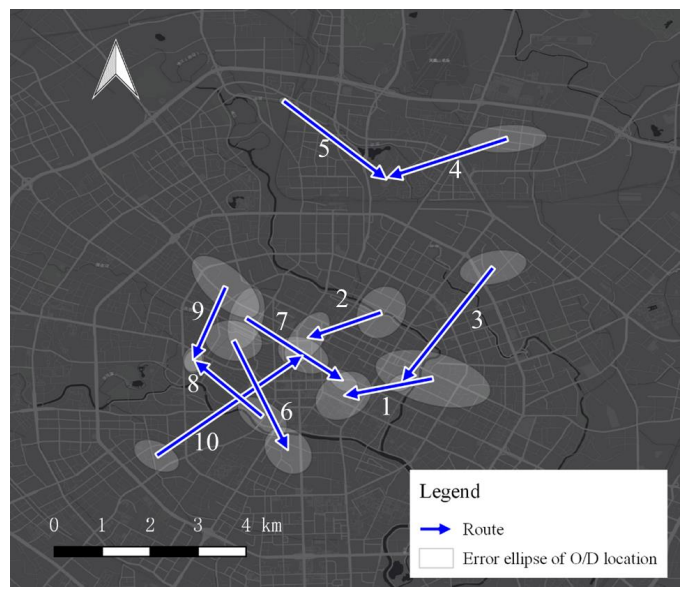

Fig. 11. Visualization of the top 10 *CGD* Clusters (Routes)

**Table 3**
Top 10 CGD clusters

| Rank | From | To | CGD |
|------|------|-----|-----|
| 1 | Chengdu 339 | Kowloon plaza | 5.88 |
| 2 | Huaxi courtyard | Platinum Building | 2.60 |
| 3 | Shallow peninsula | City No. 2 Hospital | 2.26 |
| 4 | Shangdong No.1 | Shengxianhu MTR Station | 2.17 |
| 5 | Huamanting | Shengxianhu MTR Station | 1.88 |
| 6 | Zhongchuan Building | Huaxi Internetional | 1.84 |
| 7 | Railway Research Institute | Evergrande square | 1.86 |
| 8 | Confucian temple west street | Provincial People's Hospital | 1.71 |
| 9 | Huapaifang | Provincial People's Hospital | 1.62 |
| 10 | Mingdu Garden | Downtown | 1.60 |

### 5.3.2 *Daily Stability of Commuter*

If one cluster has a steady commuter volume every day, that corresponding route is more potential for setting public transportation service. To evaluate the stability of commuters on each cluster, given the list of the daily commuter count of cluster $i$, $DCC_i = [count_i^1, count_i^2, ..., count_i^5]$, the $DSC$ is defined and calculated as (9).

$$DSC_i = \frac{\sigma(DCC_i)}{\mu(DCC_i)}$$ (9)

Where $i$ is the index of a cluster; $\sigma$ and $\mu$ respectively represent the standard deviation and mean value of $DCC$.

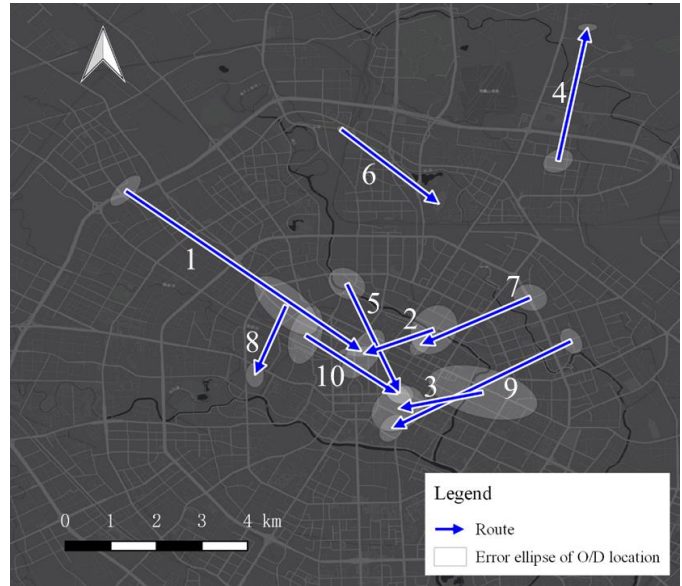

Fig. 12. Visualization of the top 10 DSC Clusters(Routes)

The cluster with a lower value of $DSC$ has a more stable commuter volume. To find the ideal routes, we sorted the value from the lowest to the highest. The top 10 clusters are shown in Fig. 12 and Table.4. From the result, the top 10 $DSC$ clusters all departed from the north of the city, reflecting the uneven development of public transportation. The cluster departed from Chadianzi to Downtown has fixed five commuters every day, making the route with the highest DSC. Some clusters (from Huaxi courtyard to Platinum building, from Chengdu 339 to Kowloon plaza, and from Railway Research Institute to Evergrande square) also have strong stability of commuters. Moreover, some clusters towards hospitals, such as from Shandong No.1 to Hospital of the Western War Zone, from Huapaifang to Provincial People`s Hospital, have a high DSC; thus, we speculate that the passengers in these routes are more likely to be medical staff.

**Table 4**
Top 10 DSC clusters

| Rank | From | To | DSC |
|------|------|------|------|
| 1 | Chadianzi | Downtown | 0 |
| 2 | Huaxi courtyard | Platinum Building | 0.104 |
| 3 | Chengdu 339 | Kowloon plaza | 0.116 |
| 4 | Shangdong No.1 | Hospital of the Western War Zone | 0.126 |
| 5 | Wuding community | Evergrande square | 0.129 |
| 6 | Huamanting | Shengxianhu MTR Station | 0.136 |
| 7 | Shallow peninsula | Shuangyanjing community | 0.156 |
| 8 | Huaxin Garden | Provincial People's Hospital | 0.158 |
| 9 | Yihe Home | Chengdu Wealth Center | 0.178 |
| 10 | Railway Research Institute | Evergrande square | 0.182 |

### 5.3.3 Concentration of Departure Times

In general, the concentration of departure time affects the potency for emerging demand-responsive public transportation services such as customized buses or carpooling. To describe the concentration degree of departure time in each cluster, $CDT$ is defined as (10). Given the collection of each trip departure time in route $i$, $T_i^O = [t_{i,1}^o, t_{i,2}^o, t_{i,3}^o, ..., t_{i,j}^o]$, $CDT$ is calculated as the information entropy(Shannon, 1948) of the distribution of $T_i^O$ :

$$CDT_i = entropy(T_i^O) \tag{10}$$

The cluster with a lower value of *CDT* has a more concentrated departure time which is more suitable to set up demand-responsive public transportation service. All clusters` *CDT* values are sorted from the lowest to the highest. The top 10 clusters are shown in Fig. 13 and Table 5. From the result, the destinations of these clusters with concentrated departure times are all large business districts, such as Evergrande square, Time impression square, Railway station, etc., indicating that people working in these business districts have regular commuting.

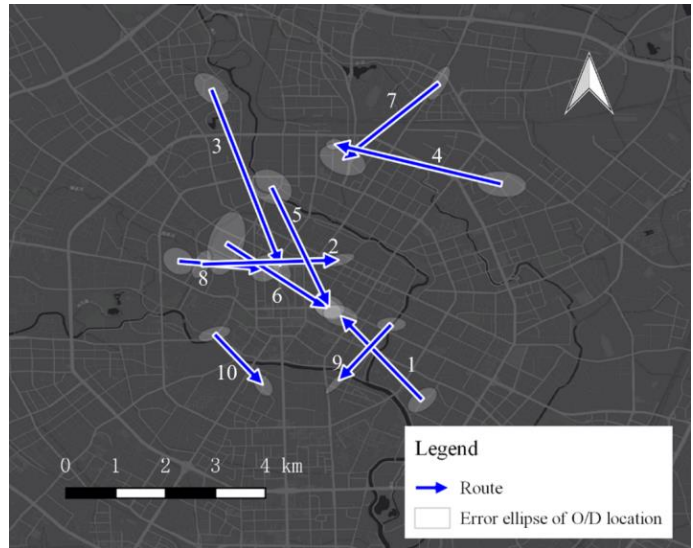

Fig. 13 Visualization of the Top 10 CDT Clusters(Routes)

**Table 5**
Top 10 CDT clusters

| Rank | From | To | CDT |
|------|------|-----|-----|
| 1 | Longhu | Evergrande square | 1.316 |
| 2 | Jujube lane | Time impression square | 1.418 |
| 3 | Jinximinyuan | Platinum building | 1.435 |
| 4 | Bali Community | Chengdu Railway Station | 1.458 |
| 5 | Baima Temple | Evergrande square | 1.522 |
| 6 | Railway Research Institute | Evergrande square | 1.527 |
| 7 | Xingyuanhuasheng | Dacheng Market | 1.548 |
| 8 | Qingyang Community | Teach science college | 1.560 |
| 9 | Tianyashi Community | Nanmen | 1.561 |
| 10 | Greatwall Community | Lianyi Building | 1.564 |

### 5.3.4 Comprehensive evaluation

Finally, to consider the above aspects simultaneously, we define *PTS-Score* by combining the three indicators to obtain a comprehensive evaluation of routes potential for setting up public transportation service such as customized bus, as (11). Fig. 14 shows the clusters sorted by the *PTS-Score*.

$$PTS\text{-}Score_i = \frac{CGD_i}{DSC_i \times CDT_i} \tag{11}$$

Further, to analyze the pattern of the clusters, the relationship of rank between the *PTS-Score* and three indicators is shown in Fig. 15, in which the routes with higher *PTS-Score* have a pattern of "higher CGD - higher DSC - lower CDT", indicating these clusters have large, daily regular commuters, and the depart time is evenly distributed in the commuting travel period. On the contrary, the routes with lower *PTS-Score* have a pattern of "lower CGD - lower DSC - higher CDT", indicating these clusters are sporadic, collective travel events, and the routes corresponding to these clusters are not suitable for setting public transportation service. Besides, more patterns within the routes need to be further revealed and analyzed.

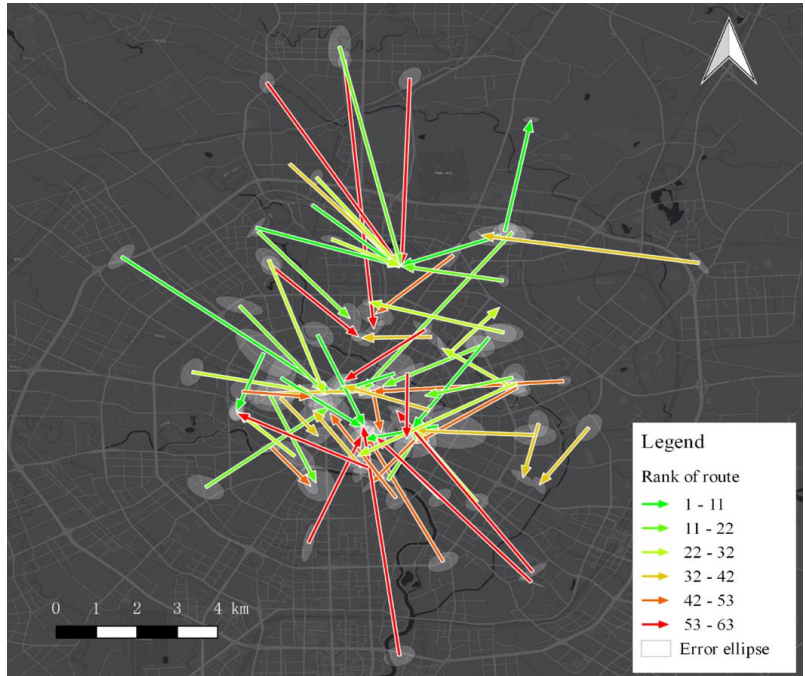

Fig. 14. The rank of routes` *PTS-Score*

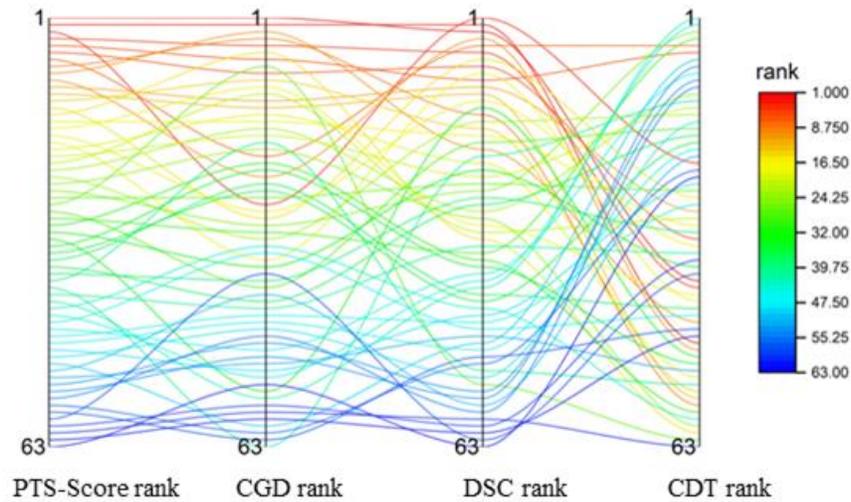

Fig. 15. The parallel plot of three evaluation indicators ranks and PTS-Score rank

## 5.4 Comparison

For comparison, the benchmark methods were adopted to detect clusters, and the experimental parameter settings were determined according to the literature. The OD flow cluster-detecting results are shown in Fig. 16.

According to Fig.16, the Density Domain Decomposition method and Local K function method only detected clusters with a fine aggregation scale (with a small O/D scope), as a result, some Top 10 routes such as the route from Bali Community and Shallow peninsula to Chengdu Railway Station are missed. The incomplete results make it difficult to paint a complete picture of urban commuting. On the contrary, though the flowHDBSCAN method identified lots of clusters, the result is hard to interpret since it is confused by low densities clusters that are not significantly clustered and noise mixed. Besides, during the study, the flowHDBSCAN is again proved to be sensitive to parameter`s value because almost all the clusters were combined into one false cluster when it increased from 11 to 12.

In conclusion, our method can identify OD flow clusters of urban travel data effectively with significantly clustered commuters and easily identifiable routes which can provide effective support for designing public transportation service, such as customized bus.

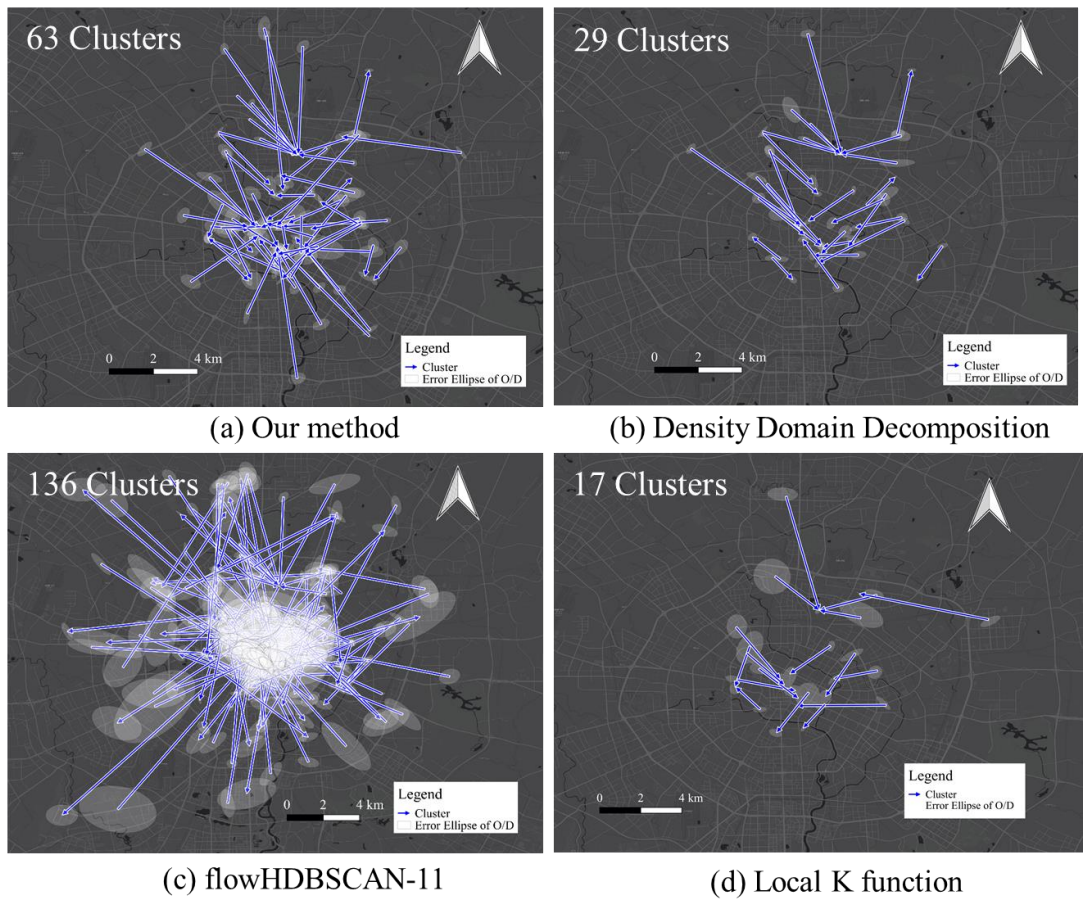

Fig. 16. Comparison of our method with benchmark methods in the study area.

## 6. Conclusion

OD flow clustering has been widely applied to reveal the dominant trend and spatial structure of urban resident travel. The existing methods for OD flow cluster-detecting are limited both in specific aggregation scale and the uncertain result due to different parameters setting, which is difficult for complicated OD flows clustering under spatial heterogeneity. To address these limitations, we proposed a novel cluster-detecting method based on the OPTICS algorithm. The proposed method adaptively determines the parameters from the dataset using Ripley`s K Function and Kullback-Leibler Divergence without manual intervention. Experiments indicated that our method outperforms the three state-of-the-art methods evaluated by Silhouette Coefficient, V-measure, and Fowlkes Mallows index for the following reasons: 1) Generating the cluster structure by OPTICS made the identified clusters with various aggregation scales and avoided the clusters being merged or ignored incorrectly; 2) The K function ensured the statistical significance of the clusters to avoid generating false clusters; 3) The LOF method denoised the results that made the depiction of mobility trends of clusters more accurate. In the case study, the proposed method is applied to identify commuter group routes and assist public transportation service design. The comparing results with benchmark methods shown that our method can obtain significant clustered commuter groups with various scales and delineate potential routes clearly.

However, in this study, the determination of MaxEps by spatial statistics requires massive Monte Carlo simulations, which requires a high computational cost. Besides, the potential for public transportation service on the detecting routes needs further evaluation and validation.

## Acknowledgment

The authors would like to sincerely thank the anonymous reviewers for their constructive comments and valuable suggestions that improved the quality of this article. This work was supported in part by the National Key Research and Development Plan of China (grant numbers 2017YFB0503604 and 2016YFE0200400) the National Natural Science Foundation of China (grant numbers 41971405, 41671442) and the Open research fund program of LIESMARS, Wuhan university (No. 19S01). The authors thank the DIDI Chuxing for providing the data used in this paper. Data source: DIDI Chuxing GAIA Initiative (https://gaia.didichuxing.com). Thanks to Dr. Chang Ren for providing inspiration for the case study. The corresponding authors of this manuscript are Prof.

Luliang Tang and Xue Yang.